\documentclass{ws-ijmpe}

\begin{document}

\markboth{H. Al~Falou, A. Leprince, N.A. Orr}{INVESTIGATION OF CONTINUUM STATES ...}

\catchline{}{}{}{}{}

\title{INVESTIGATION OF CONTINUUM STATES IN THE UNBOUND NEUTRON-RICH
N=7 AND 9 ISOTONES VIA\\
BREAKUP AND KNOCKOUT\footnote{Contribution to NIIGATA2010 - ``International Symposium 
on Forefronts of Reseach in Exotic Nuclear Structures'', Tokamachi, Japan, 1--4 mars 2010.
}}

\author{\footnotesize H.~Al~Falou\footnote{Present Address: \textsc{TRIUMF}, 4004 Wesbrook Mall,
Vancouver, BC V6T 2A3
Canada}, A.~Leprince, N.A.~Orr\footnote{Presenting author, e-mail: orr@lpccaen.in2p3.fr} \\
for the \textsc{LPC--CHARISSA--DEMON Collaboration}}

\address{LPC-ENSICAEN, IN2P3-CNRS et Universit\'e de Caen,\\
14050 Caen cedex, France}

\maketitle


\begin{abstract}

The structure of the unbound nuclei $^{9}$He, 
$^{10}$Li and $^{13}$Be has been explored using breakup and proton-knockout from
intermediate energy $^{11}$Be and $^{14,15}$B beams.  In the case of both N=7 isotones, virtual
$s-$wave strength is observed near threshold together with a  
higher-lying resonance.
A very narrow structure at threshold in the $^{12}$Be+n relative energy spectrum is demonstrated to
arise from the decay of the $^{14}$Be$^*$(2$^+$), discounting earlier reports of a 
strong virtual $s-$wave state in $^{13}$Be.

\end{abstract}

\vspace*{22pt}

The light nuclei have long provided a fertile testing ground for our understanding of
nuclear structure. From an experimental point of view, this is the only region for which nuclei
lying beyond the neutron dripline are presently accessible. Theoretically, models
incorporating explicitly the continuum are being developed. In addition, the structure of
unbound systems, such as $^{10}$Li, is key to constructing three-body descriptions of two-neutron
halo, such as $^{11}$Li, and related nuclei.

One of the approaches well adapted to the study of nuclei far from stability is that of
``knockout'' or breakup of a high-energy radioactive beam. Here we report 
on measurements using secondary beams (35 MeV/nucleon) 
of $^{11}$Be and $^{14,15}$B to investigate the low-lying level structures of $^{9}$He, 
$^{10}$Li and $^{13}$Be.  The experiments employed beams delivered by the 
\textsc{LISE3} separator at \textsc{GANIL}.  The beam velocity charged fragments and neutrons emitted in the forward 
direction from the reactions on a carbon target were identified and the 
momenta determined using a Si-Si-CsI array coupled to a large-scale neutron array.
These measurements allowed the fragment+neutron ({\it f-n}) relative energy spectra to be reconstructed.  
In order to interpret the spectra, simulations, which were validated using known resonances (such as $^{7}$He$_{g.s.}$), were developed to model the response function of the experimental
setup.  Detailed accounts of the work presented here may be found elsewhere\cite{Hicham-these,Anne-these,JLL-B16}.

\begin{figure}[th]
\centerline{\psfig{file=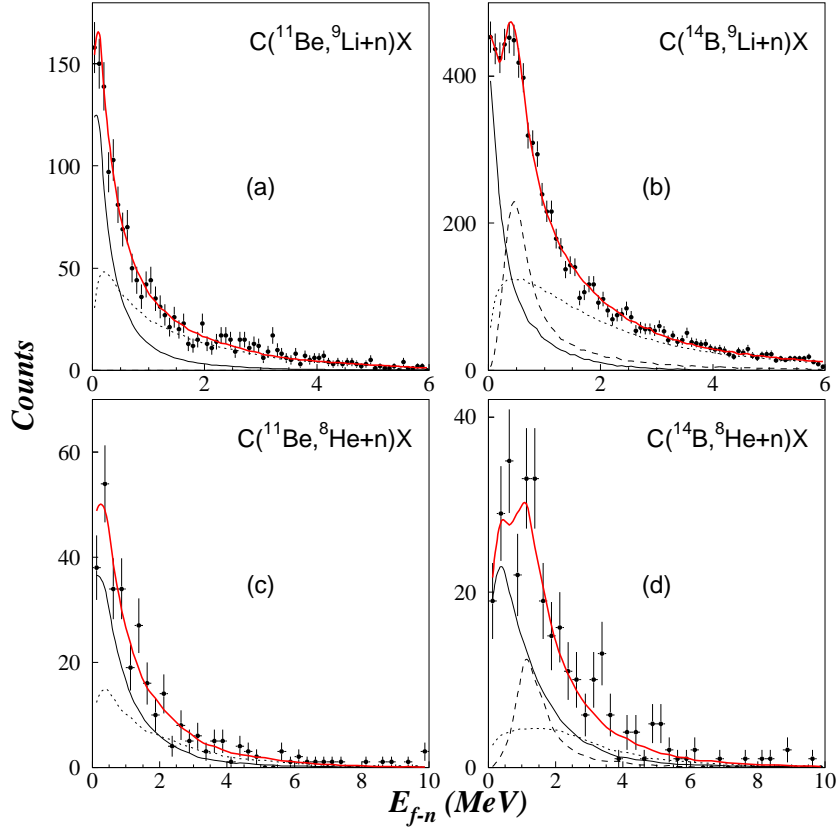,width=11cm}}
\vspace*{6pt}
\caption{Relative energy spectra for the $^{9}$Li+n and $^{8}$He+n systems for different reactions.  
The dotted lines represent the uncorrelated background.  The thin solid
lines are the virtual $s-$states, while the dashed lines are the resonances.  The thicker
solid line is the overall adjustment.}
\end{figure}

The results obtained for the $^{8}$He+n and $^{9}$Li+n systems are shown in Fig.~1.  As demonstrated in
our work on the C($^{17}$C,$^{15}$B+n) reaction\cite{JLL-B16}, the description 
of the relative energy spectra
require, in addition to discrete final states, a broad and rather featureless continuum of 
uncorrelated events which may be generated via event mixing\cite{JLL-these}.  Both spectra for
$^{9}$Li+n (Fig. 1(a) and (b)), as well as that for $^{8}$He+n derived from two-proton knockout from $^{11}$Be (Fig. 1(c)), exhibit 
significant strength just above threshold which can be most satisfactorily described by
the presence of a virtual s-wave scattering state.  The results for the C($^{11}$Be,$^{9}$Li+n) and
C($^{11}$Be,$^{8}$He+n) reactions
are in line with what may be expected
on the basis of simple considerations, whereby proton only removal from the projectile should leave the 
neutron configuration undisturbed\cite{MSU-He9}.  Given the dominant $s-$wave neutron component 
in $^{11}$Be$_{g.s.}$,
proton knockout to $^{10}$Li and $^{9}$He should populate preferentially $s-$wave final states.
In the case of $^{9}$Li+n, a scattering length around
-17~fm was deduced, whereas that for $^{8}$He+n is close to 0~fm 
($a_s$=~-3~--~0~fm at the 3-sigma level), signifying a very weak fragment-neutron interaction.  
The $^{10}$Li result is 
in line with other studies, including high-energy neutron knockout from $^{11}$Li\cite{GSI-Li10-12,GSI-Li10Be13}, whilst that for $^{9}$He, despite being in conflict with
similar work at slightly lower energies\cite{MSU-He9}, has been confirmed by a very recent
study at relativistic energies\cite{GSI-He9}. 

The $^{9}$Li+n spectrum from the breakup of 
$^{14}$B clearly displays the presence of a higher lying state some 0.5~MeV above threshold, which may be identified with the expected $p-$wave resonance observed in other studies\cite{GSI-Li10-12,GSI-Li10Be13}.
Interestingly, breakup of $^{15}$B exhibits an enhanced yield to this resonance relative to the 
$s-$state\cite{Anne-these}, as displayed in Fig. 2.
Despite suffering from limited statistics, the $^{8}$He+n relative energy spectrum obtained from 
breakup of $^{14}$B is consistent with the presence of the very weakly interacting $s-$wave 
virtual state and a resonance around 1.2~MeV above thresold\cite{Bohlen-Seth}.

\begin{figure}[th]
\centerline{\psfig{file=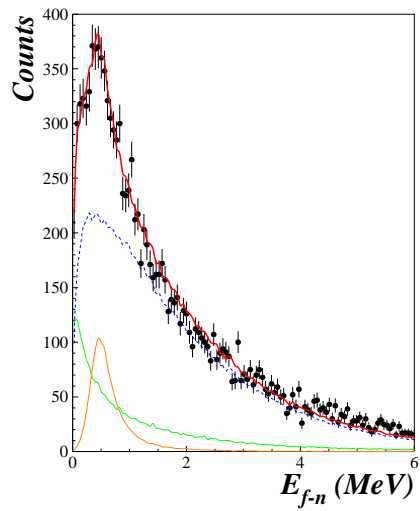,width=5.5cm}}
\vspace*{6pt}
\caption{Relative energy spectra for the $^{9}$Li+n from breakup of a $^{15}$B beam.  
The dotted lines represent the uncorrelated background, whilst the thin solid
lines are the virtual $s-$state and resonance (see text).  The thicker
solid line is the overall adjustment.}
\end{figure}

\begin{figure}
\centering
\begin{tabular}{cc}
\psfig{file=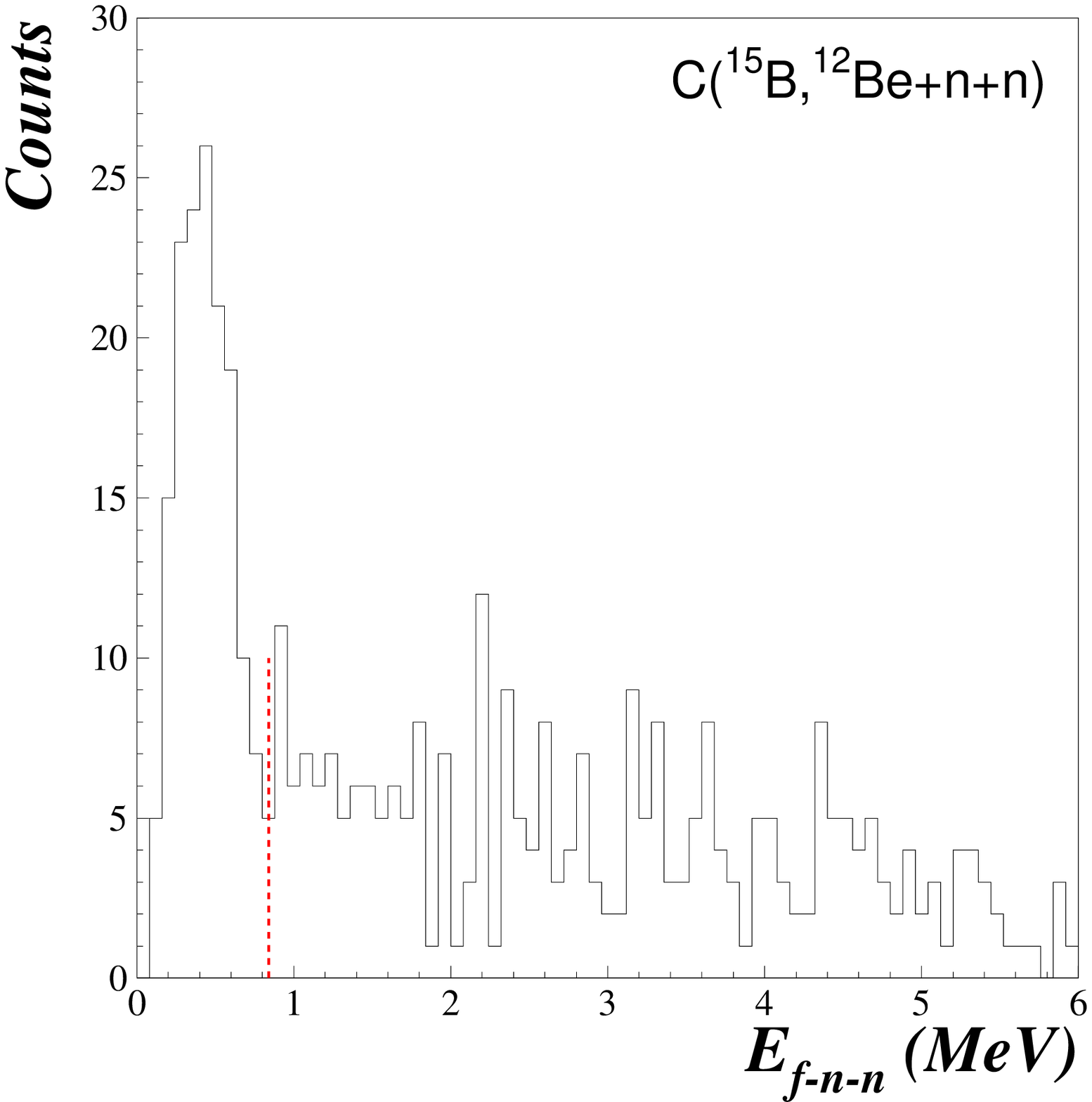,width=0.45\linewidth} &
\psfig{file=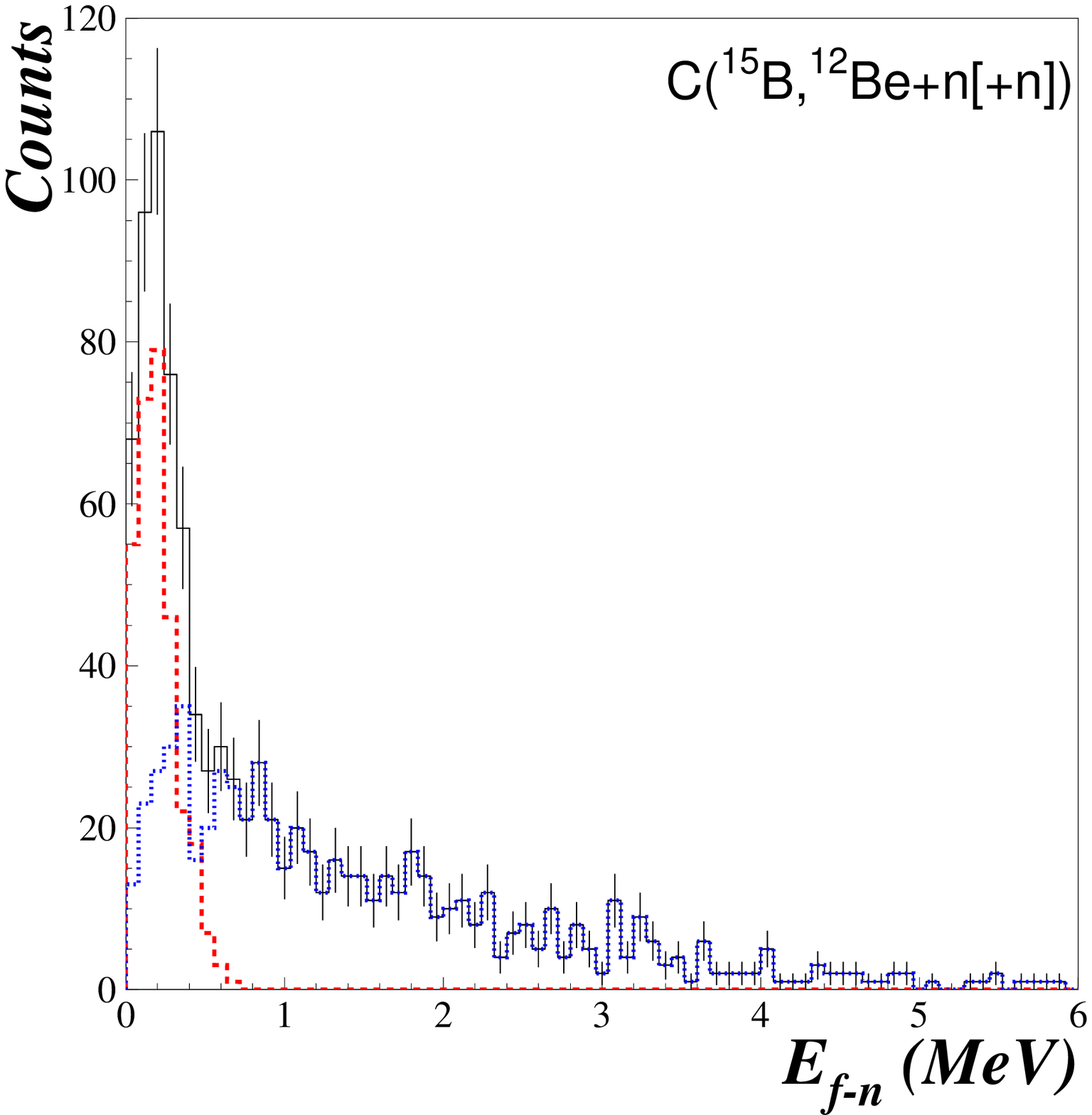,width=0.45\linewidth} 
\end{tabular}
\vspace*{6pt}
\caption{Left panel: Relative energy spectrum for the $^{12}$Be+n+n system.  
Right panel: $^{12}$Be+n relative energy spectrum 
derived from $^{12}$Be+n+n events, where events arising from the decay of $^{14}$Be$^*$(2$^+$) 
are shown as the dashed histogramme close to threshold.}
\end{figure}

Turning to the N=9 system $^{13}$Be, we have investigated an earlier report of pronounced $s-$wave virtual 
strength at threshold\cite{MSU-Be13}, which is at varience with all other studies\cite{JLL-these,GSI-Li10Be13,JLL-Few-Body,Kondo-Be13}.  The distinguishing features of the earlier
work were: a collinear geometry for the fragment and neutron detection, which limited the
range of accessible relative energies to close to threshold, and the 
fragmentation of $^{18}$O to produce the $^{12}$Be+n system.  Given that in such  
reactions (where protons and neutrons are removed) the outgoing channel of interest will tend to be 
produced via the
neutron decay of more neutron-rich unbound states, it is likely that the majority of the $^{12}$Be+n yield from $^{18}$O results from the decay of unbound states in $^{14}$Be.
Any narrow resonances in the $^{14}$Be continuum could then leave an imprint on the $^{12}$Be+n relative
energy spectrum as a narrow, resonance-like structure.

In this spirit, we have investigated the decay of $^{14}$Be$^*$ through its population in single-proton 
removal from $^{15}$B.  The resulting $^{12}$Be+n+n relative energy spectrum (Fig.~3) exhibits a 
well populated peak, some
300~keV above threshold, which corresponds to the 2$^+_1$ state 
of $^{14}$Be\cite{RIKEN-Be14-Ex}.  The $^{12}$Be+n relative energy spectrum reconstructed from these events
is also displayed in Fig.~2.  When  
the $^{12}$Be+n events are confined to those arising from the decay of $^{14}$Be$^*$(2$^+$), they are
seen, unsurprisingly, to constitute
the very narrow structure at threshold (Fig.~3, right panel).  That is to say, the threshold strength in the $^{12}$Be+n relative energy spectrum is an artefact of the limited energy available in the decay of $^{14}$Be$^*$(2$^+$).

\begin{figure}[th]
\centerline{\psfig{file=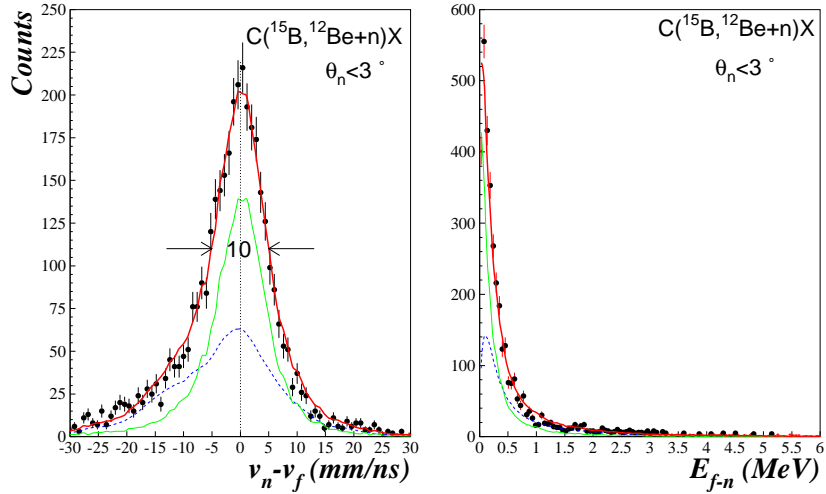,width=11cm}}
\vspace*{6pt}
\caption{$^{12}$Be+n relative velocity (left panel -- \textsc{FWHM}=10~mm/ns) and energy spectra (right panel) from the breakup of 
$^{15}$B for $\theta_n<$3$^\circ$.  The dotted line is the uncorrelated background and the thin solid line a 
virtual $s-$wave state lineshape.}
\end{figure}

We can proceed further and reproduce conditions similar to that of the $^{18}$O beam experiment by
restricting the neutron detection to very forward angles ($\theta_n<$3$^\circ$) and
analysing the $^{12}$Be+n events.  
The resultats are displayed in Fig.~4, including, as in the $^{18}$O study, 
the $^{12}$Be--n relative velocity spectrum\cite{MSU-Be13}.  Unsurprisingly, the 
restricted acceptances privilege almost exclusively the peak at threshold produced by the 
decay of $^{14}$Be$^*$(2$^+$) and result in a relative velocity spectrum identical to that
seen in the $^{18}$O beam experiment.  Moreover, both the relative velocity and energy spectra
are well reproduced by the combination of an uncorrelated background and an $s-$wave virtual state lineshape
with $a_s$~=~-10~fm as in the original 
analysis of the $^{18}$O data\cite{MSU-Be13}.
In summary, we conclude from the present analyses, that there is no evidence for a strong $s-$wave virtual state at threshold in $^{13}$Be.

\vspace*{12pt}

The authors would like to thank their colleagues in the \textsc{LPC--CHARISSA--DEMON} Collaboration and
acknowledge the excellent support provided by the technical staff of \textsc{LPC} and \textsc{GANIL}.

\end{document}